\documentclass{aa}  

\usepackage{graphicx}
\usepackage{upgreek}
  
\usepackage{txfonts}
\usepackage{color}
\newcommand{\CII}{C\,{\sc ii}}

\begin{document}
 
 \def \bmath #1 {{\hbox{\boldmath{$#1$}\unboldmath}}}

 \title{\bf Formation of interstellar SH$^+$ from vibrationally excited H$_2$:\\
 Quantum study of S$^+$ + H$_2$ $\rightleftarrows$ SH$^+$ + H reactions
 and inelastic collisions}
 \authorrunning{Zanchet et al.}
\titlerunning{Formation and destruction of SH$^+$}
 
\author{Alexandre Zanchet\inst{1}
  \and
  Francois Lique\inst{2}
  \and
  Octavio Roncero\inst{1}
  \and
  Javier R. Goicoechea\inst{1}
  \and
  Niyazi Bulut\inst{3}
}                                                
     \institute{Instituto de F{\'\i}sica Fundamental (IFF-CSIC),        
          C.S.I.C.,                                   
          Serrano 123, 28006 Madrid, Spain.
       \and
       LOMC - UMR 6294, CNRS-Universit\'e du Havre, 25 rue Philippe Lebon,
       BP 1123 - 76 063 Le Havre cedex, France
       \and
           Firat University, Department of Physics, 23169
           Elazig, Turkey
     }

\abstract{    
        The rate constants for the formation, destruction, and collisional excitation of SH$^+$ 
        are calculated from quantum mechanical approaches using 
        two new SH$_2^+$ potential energy surfaces (PESs) of $^4A''$ and $^2A''$ electronic symmetry.
        The PESs were developed to describe all adiabatic states correlating to the SH$^+$ ($^3\Sigma^-$) + H($^2S$) channel.
        The formation of SH$^+$ through the S$^+$ + H$_2$ reaction is endothermic by $\approx$ 9860 K, and requires at least
        two vibrational quanta on the H$_2$ molecule to yield significant reactivity.
        Quasi-classical calculations of the total formation rate constant for H$_2$($v=2$)
        are in very good agreement with the quantum results above 100K. Further quasi-classical
        calculations are then performed for $v=3$, 4, and 5 to cover all vibrationally
        excited H$_2$ levels significantly populated in dense photodissociation regions (PDR).
        The new calculated formation and destruction rate constants are two to six times larger than the previous ones and have been
        introduced in the Meudon PDR code to simulate
        the physical and illuminating conditions in the Orion bar prototypical PDR.
        New astrochemical models based on the new molecular data produce four times larger
        SH$^+$ column densities, in agreement with those inferred from recent ALMA observations of the Orion bar.
}

\keywords{Astrochemistry - Molecular data — Molecular processes --- ISM: photon-dominated region (PDR)  – ISM: clouds}
  \maketitle


\section{Introduction} \label{sec:introduction}

Molecular hydrogen is by far the most abundant molecule in the Universe and triggers the chemistry in the interstellar medium (ISM) through reactions with 
the most abundant atoms and ions.
Owing to their high reactivity, the formation of molecular hydride cations 
is an important first step toward the synthesis of
more complex interstellar molecules.

In UV-illuminated environments such as the low-density diffuse clouds
or the edges of dense molecular clouds close to massive stars, the
so-called photodissociation regions \citep[PDRs,][]{Hollenbach:97}, the formation of hydride cations can start through the following reaction. 
\begin{eqnarray}\label{hidrogenation-reaction}
     {\rm M}^+ + {\rm H}_2 \rightarrow {\rm MH}^+ + {\rm H},
\end{eqnarray}
where  M$^+$ is an ion typically formed by photo-ionization of element M,
such as carbon or sulfur, with an ionization potential below 13.6\,eV.

The launch of the \textit{Herschel} satellite in 2009 opened a new frequency window in the far-IR/submillimeter wavelengths 
allowing for the detection of molecular hydrides in the ISM \citep[for a review, see][]{Gerin:16}.
The \textit{Herschel} satellite allowed for the detection of
rotational line emission from CH$^+$ or SH$^+$ 
in dense interstellar PDRs \citep{Nagy-etal:13,Naylor:10,Pirelli-etal:14,Joblin-etal:18}, in the irradiated walls of protostellar outflows \citep{Falgarone:10,Benz-etal:10,Benz:16},
and in the circumstellar envelopes around hot planetary nebulae 
\citep{daSilva:18}. 

For CH$^+$ and SH$^+$ ions, \mbox{reaction~(\ref{hidrogenation-reaction})} is very endothermic
($\Delta E/k=4300$ and 9860~K, respectively) and the above interstellar detections may seem surprising. In UV-irradiated environments however, H$_2$ can be radiatively pumped to vibrationally excited levels \citep[e.g.,][]{Black:76}. These vibrational states have internal energies high enough to overcome reaction
endothermicities \citep{Sternberg:1995}. Indeed, laboratory experiments showed that for
CH$^+$, \mbox{reaction~(\ref{hidrogenation-reaction})} becomes exothermic and fast
if H$_2$\,($v\ge 1$) \citep{Hierl1997}. \mbox{State-to-state} rate constants
for this reaction have have been calculated from quantum calculations  \citep{Zanchet-etal:13} and PDR and excitation models using these data predict CH$^+$ abundances and rotational line intensities close to the observed ones \citep[e.g.,][]{Agundez-etal:10,Godard-Cernicharo:12,Nagy-etal:13,Faure-etal:17,Joblin-etal:18}. In addition, mapping observations of the Orion molecular cloud have revealed  spatial correlation
between the [\CII]\,158\,$\mu$m emission (from ionized carbon C$^+$), the IR H$_2$ ($\geq$1) emission, and the CH$^+$ ($j=1-0$) rotational emission \citep{Goicoechea:19}. This work observationally demonstrates that the interstellar CH$^+$ emission is widespread in \mbox{UV-irradiated} dense gas, and that its main formation route is \mbox{reaction~(\ref{hidrogenation-reaction})}. 

Formation of SH$^+$  through  \mbox{reaction~(\ref{hidrogenation-reaction})}
is much more endothermic than that of CH$^+$ and 
 reaction (\ref{hidrogenation-reaction})
only becomes
exothermic when H$_2$\,($v\geq2$) \citep[e.g.,][]{Zanchet-etal:13b}.
{{ This is true considering H$_2$ vibrational levels alone. However, we note that taking into account
    H$_2$ rovibrational levels, reaction (1) becomes exothermic for v=0, j$\ge$11 and  v=1,
    j$\ge$7.
These rotationally excited levels within a given vibrational state can be populated,
but their observed column densities are much lower than those of the low-j levels of vibrationally excited H$_2$
(e.g., \cite{Habart-etal:11,Kaplan-etal:17}).  }}.
Still, SH$^+$
has been detected in low-density diffuse clouds, 
for example through \mbox{{absorption}} lines measurements \citep{Menten-etal:11,Godard-etal:12}
and in denser and more strongly \mbox{UV-irradiated} environments
through \mbox{{emission}} lines \citep{Benz-etal:10, Nagy-etal:13}.
Contrary to CH$^+$, SH$^+$ can be observed from ground-based millimeter-wave telescopes \citep[][]{Muller-etal:14,Halfen:15}. Then, high-angular interferometric images
of the Orion Bar PDR  obtained with ALMA have revealed that the SH$^+$ emission nicely delineates the H$_2$ dissociation front at the edge of the PDR \citep[][]{Goicoechea-etal:17}, the same gas layers where H$_2$ molecules are 
\mbox{UV-pumped} to highly excited vibrationally levels \citep[][]{Kaplan-etal:17}.
These results further confirm the need for accurate state-to-state
reaction rates for reaction (\ref{hidrogenation-reaction}).

In addition, the starting hydrogen abstraction reactions involving H$_2$ and S, S$^+$, and SH$^+$ are all highly endothermic \citep{Neufeld-etal:15} and this implies that SH$^+$ is destroyed in reactions with atomic hydrogen and
electron recombinations. Both H and $e^-$ are abundant in UV-irradiated gas, and therefore SH$^+$ is expected to be relatively reactive and a short-lived hydride in PDRs. 

Therefore, it is very relevant to study the first steps of sulfur chemistry.
In particular, \citet{Goicoechea-etal:17} concluded that the abundances of SH$^+$
inferred from the ALMA images of the Orion Bar were $\sim$ 3-30 times higher than those predicted by a state-of-the-art PDR model using specific rate constants, computed by some of us for the 
S$^+$ + H$_2$ ($v$) reaction (with $v$ from 0 to 4) using the ground-quartet-state potential energy surface (PES) and a quasi-classical trajectories method \citep{Zanchet-etal:13b}.

In this work, we focus on a \textit{ab initio} quantum study of the relevant rate constants for the formation, destruction, and excitation of SH$^+$. By studying the reaction dynamics using a quasi-classical trajectory (QCT)
method on a new PES, \cite{Zanchet-etal:13b} demonstrate that the S$^+$($^4S$) + H$_2$(v$\ge$ 2) collisions was an efficient way to produce SH$^+$. The PES was also used to perform a quantum study \citep{Zanchet-etal:16} that showed that at low collisional energies some resonances appear, associated to roaming, and that the quantum reaction cross section for H$_2(v=2,j=0)$  was considerably larger than the classical ones. These differences disappeared for higher rotational excitation of H$_2$. 
The same PES was also used to estimate the destruction rate constants of SH$^+$ colliding with H. However, this rate of destruction can be considered as not very accurate since the PES was designed essentially to study the formation.

More recently, a new PES has been calculated using a larger basis set in the {\it ab initio} calculations \citep{Song-etal:18}. From this PES, these latter authors derived slightly larger QCT cross sections than those previously reported.
These latter works show the necessity for high-level  {~\it ab initio} calculations to correctly describe the interactions between the three atoms and to take in account quantum effects in the reaction dynamics in order to get accurate estimations of the rate constants. 

\begin{figure}[h]
\begin{center}
\includegraphics[width=7.cm]{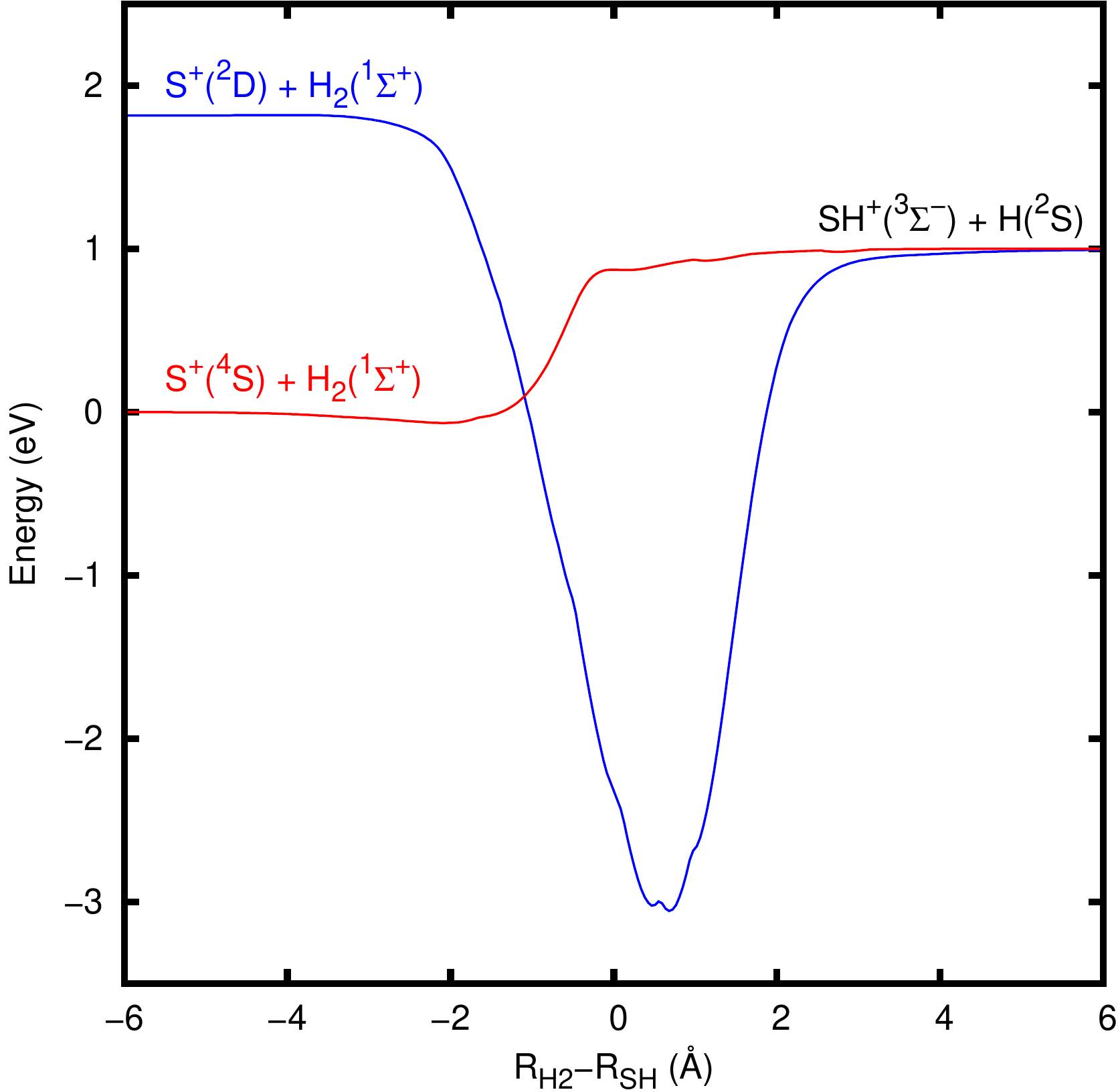}

\caption{\label{meps}{
    Minimum-energy path for the S$^+$+H$_2$
    $\rightarrow$ SH$^+(^3\Sigma^-)$ + H($^2S$) reaction
    considering the quartet and doublet states correlating
    to the SH$^+(^3\Sigma^-)$ asymptote.
    {The abscissa is the reaction coordinate, defined
  as the difference of
  the H$_2$ and SH distances, $R_{H_2}$ and $R_{SH}$, respectively,
    in \AA. The potential energies displayed correspond
    to the minimum of energy in the remaining two internal coordinates,
    $i.e.,$ the angle between the H-H and S-H bonds and the
    coordinate defined as $(R_{H_2}+R_{SH})/2$. 
     }{ At distances of 6\AA, the interaction energy is lower than 1 meV, and cannot be appreciated
in the figure. }
}}
\end{center}
\end{figure}

In order to describe the reaction
\begin{eqnarray}\label{destruction-reaction}
  SH^+(^3\Sigma^-) + H(^2S) &\rightarrow&  S^+ + H_2  \quad\quad {\rm destruction,\quad 2a)} \nonumber\\
   &\rightarrow&  H + SH^+  \quad\quad {\rm exchange,\quad 2b),}\nonumber
\end{eqnarray}
not only is the quartet PES needed, but also a doublet PES, since the spins of both SH$^+$($^2\Sigma^-$) and H($^2 S$)
can give rise to both doublet and quartet states which are asymptotically degenerate, as shown in Fig.~\ref{meps}. 
In this work the objective is to present a new set of PESs for the quartet and doublet states  using a larger electronic basis set than
that used by \cite{Zanchet-etal:13b} and \cite{Song-etal:18}.{It should be noted
that during the preparation of this work, a new PES for the doublet
electronic state was also published \citep{Zhang-etal:18} and that QCT calculations
were performed for the formation from the excited state of
S$^+(^2D)${. Reactions of H$_2$ molecules with S$^+$ ions
  in electronic excited states are expected to be negligible
  inside molecular clouds  and  only relevant at the PDR/HII
  interface layers where a small fraction of H$_2$ might exist.}
}

The paper is organized as follows. 
First the construction of the new PESs for the two electronic states (quartet and doublet) is described in Sect. 2.
Sections 3 and 4 are devoted to the reaction dynamics on the quartet and doublet
states, respectively.
Finally, in Sect. 5, the astrophysical implications of the new results obtained for both the formation and destruction of SH$^+$ are discussed.

\section{{\it Ab initio} calculations and analytical fit}

The $^4A''$  and $^2A''$ electronic states of the SH$_2^+$ system both correlate with the SH$^+$ ($^3\Sigma^-$) + H($^2S$) asymptote where they are degenerate. 
In the other asymptotic channel, each state correlates to a different electronic state of the S$^+$ cation,
the quartet state connects to the ground state  S$^+$($^4S$) + H$_2$($^1\Sigma^+$) while the doublet connects to the excited state  S$^+$($^2D$) + H$_2$($^1\Sigma^+$) reactants.
The $^4S$ and $^2D$ states of S$^+$ are separated by $\approx$ 1.8 eV as shown in Fig.~\ref{meps}. 
These states are only coupled by the spin-orbit coupling which is neglected in this work. 
In the  S$^+$($^2D$) asymptote,  five electronic states are degenerate but only one correlates  to the ground state of  SH$^+$ while the  four others correlate to excited states of SH$^+$ products, and hence their PESs  are not considered in this work. 
However, to get an accurate and homogeneous description of the doublet and quartet PESs in all the configuration space and in particular in the asymptotic channels,
the five doublet states as well as the quartet state were taken into account in the {\it ab initio} calculations. 
The state-average complete active space (SA-CASSCF) method \citep{Werner-knowles:85} was employed to calculate the first $^4A''$ together
with the two first $^2A'$ and the three first $^2A''$  electronic states. 
The active space considered consists in seven electrons distributed in six orbitals (5 - 9{\it a'} and 2{\it a''}) in order to include all valence orbitals
of sulfur and the 1{\it s} orbitals from both hydrogen atoms. 
The obtained state-average orbitals and multireference configurations were then used to calculate both the lowest $^4A''$
and the lowest  $^2A''$ state energies with the internally contracted multireference configuration interaction method (ic-MRCI)
including simple and double excitation \citep{Werner-etal:88} and Davidson correction \citep{Davidson:75}. 
The 1{\it s} orbital of sulfur was kept frozen. 
For both sulfur and hydrogen atoms, the aug-cc-pV5Z basis set  (aV5Z) was used, including  {\it spdfgh} and {\it spdfg} basis functions and 
all calculations were done using the MOLPRO suite of programs \citep{MOLPRO}.

{\it Ab initio} calculations were performed over 3800 geometries  of the SH$_2^+$ system.
To sample the geometries, three sets of coordinates were used.
Jacobi coordinates associated to the S$^+$ + H$_2$ channel were used for a good description of this asymptotic channel while internal coordinates H-H-S$^+$ and H-S$^+$-H were employed to sample the channel associated to SH$^+$+H.
These icMRCI+Q energies for the electronic states $^4A''$ and $^2A''$ were then fitted separately
using the GFIT3C procedure \citep{Aguado-Paniagua:92,Aguado-etal:93,Aguado-etal:98}, in which  
 a global PES is represented by a many-body expansion:
$$
V_{ABC} = \sum_A V_A^{(1)} + \sum_{AB} V_{AB}^{(2)} (r_{AB}) + V_{ABC}^{(3)} (r_{AB},r_{AC},r_{BC}),
$$
where $V_A^{(1)}$ represents the energy of the atom A ($A$ = S$^+$, H, H) in its corresponding  electronic state,
$V_{AB}^{(2)}$ the diatomic terms ($AB$ = SH$^+$, SH$^+$, HH) in the corresponding electronic state, and $V_{ABC}^{(3)}$ the three-body term($ABC$ = SHH$^+$).

The diatomic terms are written as a sum of short- and long-range contributions. 
The short-range potential is defined as a shielded Coulomb potential, whereas the long-range term is a linear
combination of modified Rydberg functions \citep{Rydberg:31}  defined as :
$$
\rho_{AB} (r_{AB}) = r_{AB} e^{- \beta_{AB}^{(2)} r_{AB} }, \qquad AB=SH^+,SH^+, HH,
$$
with $\beta_{AB}^{(2)} > 0$.
The root-mean-square (rms) error of the fitted SH$^+$ potential which is
common for both PESs since they share the same SH$^+$+H asymptote is  $\approx 0.045$ kcal/mol.
The rms of the fitted H$_2$  potential is $\approx  0.029$ kcal/mol for the quartet state where H$_2$ is the pure diatomic
and $\approx  0.346$ kcal/mol for the doublet state where the H$_2$ implicitly considers the avoided crossing arising
from the crossing of the two SH$^+$ electronic states correlating to S$^+$($^4S$) and S$^+$($^2D$).

The three-body term is expressed as an expansion:
$$
V_{ABC}^{(3)} (r_{AB},r_{AC},r_{BC}) = \sum _{ijk}^K d_{ijk} \rho_{AB}^i \rho_{AC}^j \rho_{BC}^k ,
$$
where  $\rho_{AB}=r_{AB} \exp{- \beta_{AB} r_{AB} }$ are modified  Rydberg functions \citep{Rydberg:31,Aguado-Paniagua:92}.
 For  SHH$^+$, there are only two nonlinear parameters, $ \beta_{SH^+} $ and $\beta_{HH}$,
and additional constraints in the linear parameters $d_{ijk}$ to ensure symmetry of the PES with respect
to the permutation of the two H atoms~ \citep{Aguado-Paniagua:92,Aguado-etal:93,Aguado-etal:98}. 
The linear parameters $d_{ijk}, (i+j+k)$ and the two nonlinear parameters $\beta_{SH^+}$ and $\beta_{HH}$,
are determined by fitting the  approximately 3000 calculated {\it ab initio} energies after the substraction
of the one- and two-body contributions. 
In the present case, the order  $L$ is 10 for both states, giving an overall rms error of  0.430 kcal/mol
and 0.392  kcal/mol   for the $^4A''$ and $^2A''$ states, respectively.

The two PESs exhibit completely different topographies.
The $^4A''$ state does not present any minimum out of the van der Waals wells in the asymptotic
channels and does not present any barrier to reaction.
The SH$^+$+H $\rightarrow$ S$^+$+H$_2$ reaction is exothermic on this surface and
reactive collisions are likely to occur in competition with the inelastic collisions.
On the other hand, the $^2A''$ state presents a deep insertion HSH well and does not present any barrier either.
For this state, in contrast with the previous case, the SH$^+$+H $\rightarrow$ S$^+$+H$_2$ is endothermic and only inelastic collisions can occur (pure or involving H exchange).
The main features are summarized in the minimum energy path shown in Fig.~\ref{meps} 

The present PES for the quartet state, calculated with the aV5Z basis set,
is very similar to that of \cite{Zanchet-etal:13b}, calculated with the smaller aVQZ basis set.
Two relevant differences can be appreciated.

{
The first one is the D$_e$ well depth of the SH$^+$ diatomic, which is 
deeper by 5 meV for the present, larger aV5Z basis set as compared to 
that obtained with the AVQZ basis set  \citep{Zanchet-etal:13b}, as can be 
seen in Fig. 2. In Table \ref{diatCSTE} the equilibrium distance  and  dissociation 
  energies,  D$_e$ and  D$_0$,  of  SH$^+$ obtained in previous works 
are compared with the present results. All the theoretical values of 
D$_0$ \citep{McMillan-etal:16,Stancil-etal:00,Song-etal:18,Zanchet-etal:13b} are within the experimental 
uncertainty \citep{Herzberg-etal:79,Dunlavey-etal:79,Rostas-etal:84}.
Assuming that the accuracy of theoretical calculations improves 
with the size of the basis set, we conclude that the SH$^+$ diatomic 
considered in this work is slightly more accurate than the one used 
previously \citep{Zanchet-etal:13b}. This sensitivity to the basis set 
highlights the difficulty in treating the electronic structure of third 
raw atoms, which remain to be challenging calculations.
{ On the other hand, the constants of the H$_2$ diatomic are similar using 
the AVQZ  or the AV6Z basis set,
leading to the same D$_e$. As a consequence, the increase of} the SH$^+$ dissociation 
energy yields a reduction  in  the  endothermicity of the reaction  by 
  about  5meV in our present PES, but this value may still be 
overestimated considering that \cite{McMillan-etal:16} find a D$_e$ that is 12 
meV deeper using a aV6Z basis set in a study centered on the SH$^+$ 
diatomic.

\begin{table}[bht]
\begin{center}
  \caption{\label{diatCSTE}Diatomic constants of SH$^+$ { determined theoretically (top)
  and experimentally.}}

\begin{tabular}{cccc}
\hline
\hline
r$_e$(\AA)   &   D$_e$(eV)       &    D$_0$(eV)    &   Reference\\
\hline
1.365           &     3.72         &       3.56      &  This work \\        
1.365           &     3.67         &       3.52      & \cite{Zanchet-etal:13b}\\
1.365          &     3.68         &                 & \cite{Song-etal:18}\\
1.365           &     3.52         &       3.36      & \cite{Stancil-etal:00} \\
1.354          &     3.84         &       3.68      & \cite{McMillan-etal:16}\\
\hline
1.364          &  3.65 $\pm$ 0.13 &                 &\cite{Herzberg-etal:79}\\
1.364          &                  &       3.48      & \cite{Dunlavey-etal:79}\\
1.363          &     3.70          &       3.54       & \cite{Rostas-etal:84}\\
\hline
\end{tabular}
\end{center}
\end{table}

}

The second difference is not related to the size of the basis set, but to the sampling of {\it ab initio} points on the PES.
{ Indeed, in the PES of \cite{Zanchet-etal:13b},
 a poor density of {\it ab initio} points was included
in the fit of the SH$^+$+H channel.}
As a consequence, the barrier present in the collinear configuration H-SH$^+$ was not
sampled and is not present in the former PES while it is correctly described by
the new potential.
As we see in the following sections, these differences have significant implications
for the formation and destruction of SH$^+$.
The barrier considerably reduces the acceptance cone of the SH$^+$ + H reaction as it only allows
destruction at low temperatures if the H atom collides on the hydrogen side of SH$^+$.
To put it in another way, the absence of the barrier in the potential of \cite{Zanchet-etal:13b} artificially increases
the probability of destruction allowing collisions on the S side to react when they are not supposed to do so.

\section{Collision dynamics}

\subsection{Dynamics on the quartet state}

\begin{figure}[h]

\begin{center}
\includegraphics[width=7.cm]{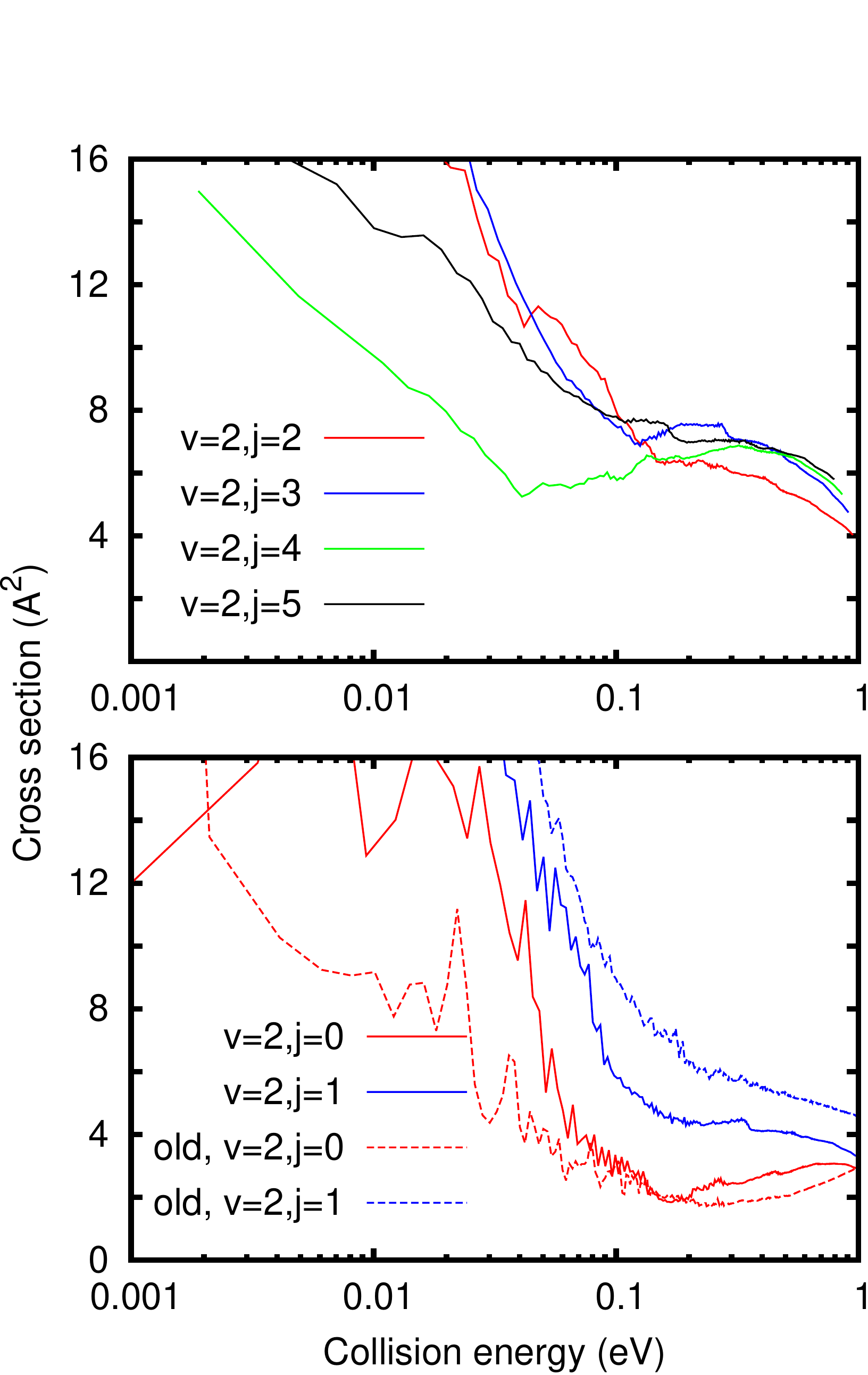}

\caption{\label{TI-quartet-formation}{
    Formation cross section for the reaction S$^+(^4S^o)$ + H$_2(v,j)$ $\rightarrow$ SH$^+(^3\Sigma^-)$ + H
     for several rotational states and $v$=2 as a function
    of collision energy, in logarithmic scale. Results marked ``old''  (dashed lines)
    are from Ref.~\cite{Zanchet-etal:16}.
}}
\end{center}
\end{figure}

The reaction dynamics on the quartet state has been studied from a time-independent treatment
based on hyperspherical coordinates. The calculations were performed with the ABC code \citep{Skouteris-etal:00}
using the parameters listed in Table ~\ref{table-ABCparameter}. 
The formation of SH$^+$ through the S$(^4S)^+$ + H$_2(v,j)$ $\rightarrow$ SH$^+(^3\Sigma^-)$ + H
reaction presents some similar trends to that obtained previously \citep{Zanchet-etal:13b,Zanchet-etal:16},
but significant quantitative difference are found.
The total reaction cross sections are shown in Fig.~\ref{TI-quartet-formation}
and compared with quantum wave packet calculations of \cite{Zanchet-etal:16}
using the previous quartet state PES of \cite{Zanchet-etal:13b}.
The present formation cross sections for collisions of S$(^4S)^+$ + H$_2$($v=2,j=0$)
is larger than that previously obtained, 
while the reactivity for the S$(^4S)^+$ + H$_2$($v=2,j=1$) reaction is the reverse. 
The cross section for collisions with H$_2$($v=2,j=1$) remains however considerably larger
than that for H$_2$($v=2,j=0$); that is, more than a factor 4 at 100 meV.
This {\it ortho}/{\it para}-H$_2$ differentiation in the reactivity is also valid
for $j$ = 2 and 3, but disappears for higher $j$.
At very low collisional energy, the cross sections on the present PES are always larger.
{ When performing the full-dimensional dynamic calculation
  it is difficult to attribute this increase to a particular detail
  of the PES. We tentatively attribute this increase to the decrease
  of the endothermicity in the present PES as compared to the previous one.
  }

In any case, the overall behavior is rather similar, 
and in both PESs, the resonances associated to roaming mediate the
reactivity at low collisional energies \citep{Zanchet-etal:16}.

        \begin{table*}
                \centering
                \caption{\label{table-ABCparameter}Parameters used in the {\scshape abc} calculations.}
                \begin{tabular}{lcr}
                        \hline \hline 
                            {\itshape jtot} & 0--100 & Total angular momentum quantum number $J$. \\
                        {\itshape jmax} & 24 & Maximum rotational quantum number of any channel. \\
                        {\itshape kmax} & 6 & Helicity truncation parameter. \\ 
                        {\itshape rmax} & 15.9 & Maximum hyper-radius $\rho_{\text{max}}$ (in \AA). \\
                        {\itshape mtr} & 600 & Number of log derivative propagation sectors. \\ 
                        {\itshape emax} & 3.5 & Maximum internal energy in any channel (in eV). \\
                        {\itshape ipar} & -1, 1 & Triatomic parity eigenvalue $P$. \\ 
                        {\itshape jpar} & -1, 1 & Diatomic parity eigenvalue $p$.
                        \\ 
                        \hline 
                \end{tabular} 
        \end{table*}

\begin{figure}[t]

\begin{center}
\includegraphics[width=7.cm]{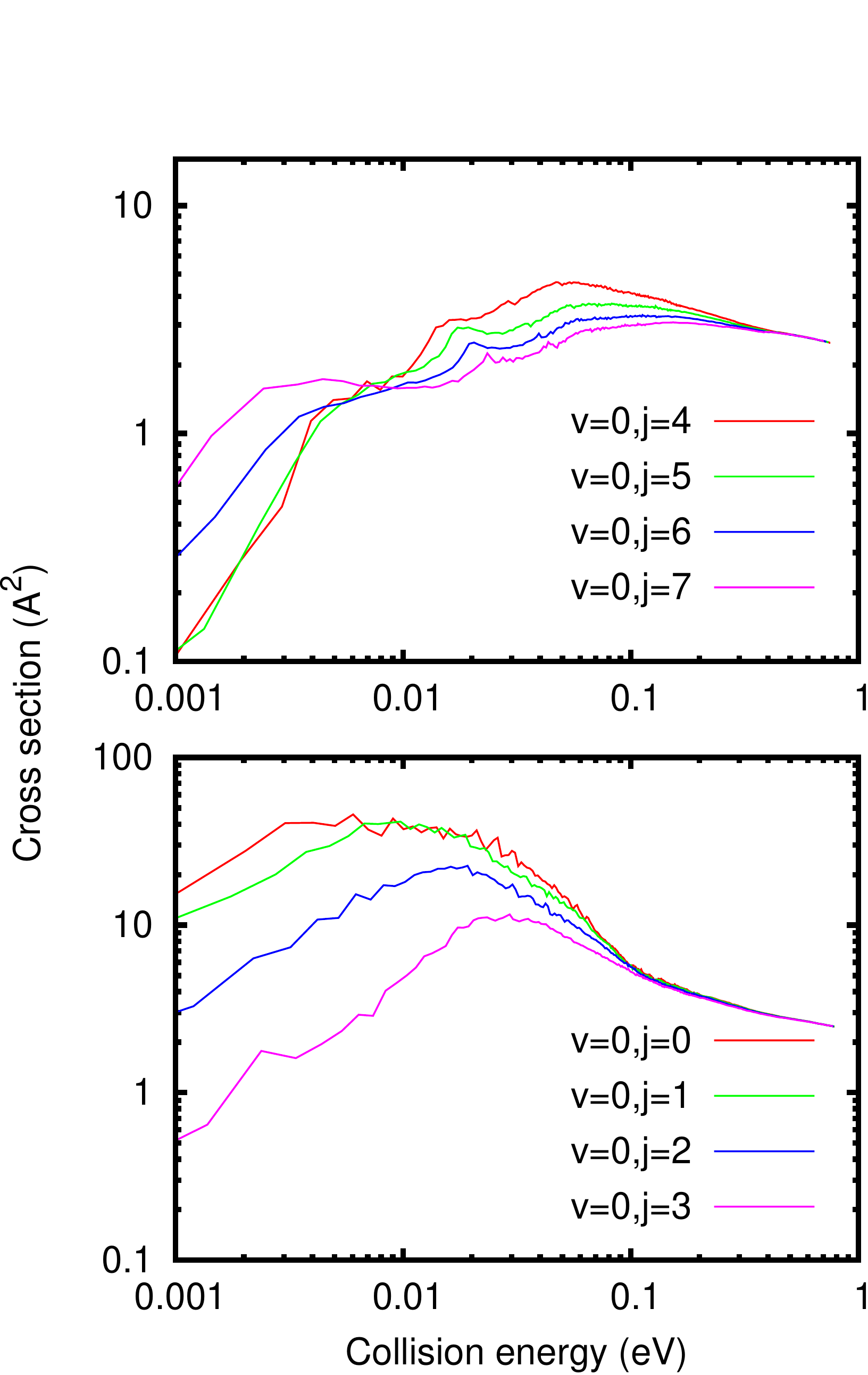}

\caption{\label{TI-quartet-destruction}{
    Destruction cross section for the reaction  SH$^+(^3\Sigma^-, v,j)$ + H $\rightarrow$
    S$^+(^4S^o)$ + H$_2$
     for several rotational states and $v$=0 as a function of collision energy. 
}}
\end{center}
\end{figure}

The destruction cross sections, for the SH$^+(^3\Sigma^-, v,j)$ + H $\rightarrow$ S$(^4S)$ + H$_2$
reaction, are shown in Fig.~\ref{TI-quartet-destruction}.
At low translational energy ($E< 0.1$ eV), the cross sections decrease with increasing
rotation, that is, the rotational excitation inhibits the reaction.
However, for $E> 0.2$ eV, all the cross sections exhibit an almost constant
value of 3-4 $\AA^2$. The order of magnitude of the formation and destruction cross sections
is very similar for high energies and nearly independent of rotational excitation of the reactants.

\begin{figure}[t]

\begin{center}
\includegraphics[width=9cm]{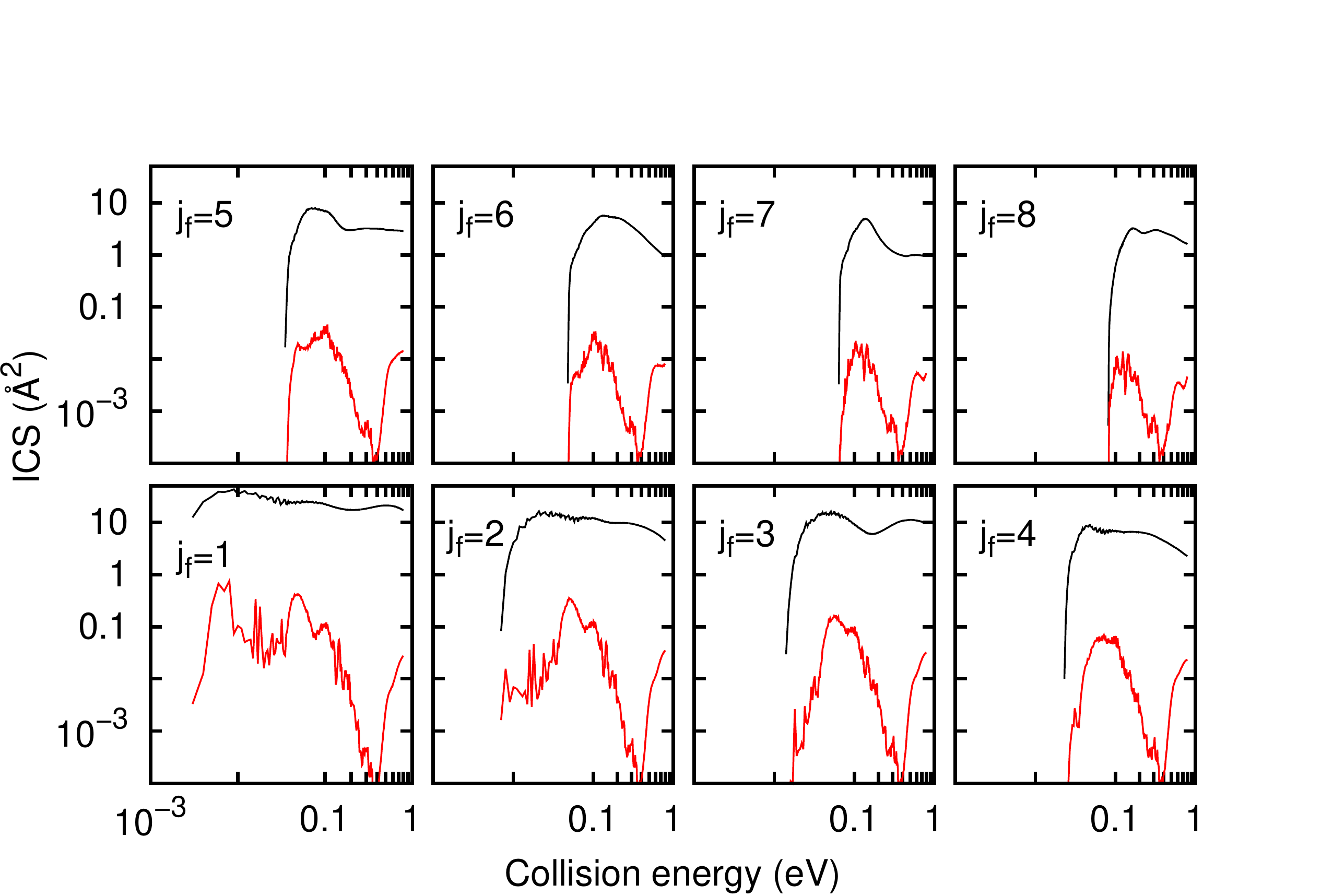}

\caption{\label{TI-quartet-inelastic}{
    Inelastic (black, solid line) and exchange (red, dashed line) cross sections
    for the collision   SH$^+(^3\Sigma^-, v=0,j=0)$ + H $\rightarrow$  SH$^+(^3\Sigma^-, v=0,j_f)$ + H
    in the quartet state
    as a function of collision energy. 
}}
\end{center}
\end{figure}

The inelastic and exchange cross sections for the H + SH$^+$(v=0,j=0)  are shown
in Fig.~\ref{TI-quartet-inelastic}. The exchange cross sections are more than two orders
of magnitude smaller than the inelastic ones, what is explained by the fact that the
collision proceeds, in this case directly, without forming a collision complex, 
making the H-exchange very improbable.
The inelastic cross sections
decrease with increasing $\Delta j$ as usual.

\subsection{Dynamics on the doublet state: inelastic and exchange}

In order to fully describe the inelastic collision, the dynamics on the doublet state
must be considered, as it accounts for one third (or 2/6) in the electronic partition function.
This state correlates with the excited sulfur cation, S$^+(^2D)$, which is nearly 2 eV higher than 
the S$^+(^4S)$ and $\approx$ 1 eV above the SH$^+$ (see Fig.~\ref{meps}).
Therefore, at low to intermediate collision energies, only the inelastic and exchange
process are possible in this electronic state, according to
Eq.~(\ref{destruction-reaction}.b) if the Born-Oppenheimer approximation is assumed. However,
the presence of the deep well of $\approx$ 4eV from the  SH$^+$ + H asymptote
makes it difficult to apply the ABC code to study this collision. Instead,
here we used a quantum wavepacket MADWAVE3 code \citep{Gomez-Carrasco-Roncero:06,Zanchet-etal:09b}
as it was applied to study the OH$^+$+H dynamics in the doublet state \citep{Bulut-etal:15b} with
a very similar deep well. The parameters used in these calculations are listed in Table \ref{table-wp-doublet}.
The state-to-state reaction probabilities have been calculated for $J=0,10,20,30, ..., 100$ and for intermediate values of
$J$; they are interpolated using the $J$-shifting approximation as done previously
\citep{Zanchet-etal:13,Gomez-Carrasco-etal:14,Bulut-etal:15b}.
After summing over all partial waves, the state-to-state cross sections were obtained.

\begin{table*}
\begin{center}
\caption{\label{table-wp-doublet}Parameters used in the wave packet calculations for the doublet PES}
\centering
\begin{tabular}{c c c c}
\hline\hline
Reactant scattering coordinate range: & $R_{min}$=0.001\AA; & $R_{max}$=40.0\AA &\\
Number of grids points in $R$:  & 600 &  & \\
Diatomic coordinate range: & $r_{min}$=0.001\AA & $r_{max}$=32.0\AA  &\\
Number of grid points in $r$: & 480 &   \\
Number of angular basis functions: & 120 &   &\\
Number of projection of total angular momentum: &$\Omega_{max}$=29 & &  \\
Center of initial wave packet:   & 19.0\AA &  &  \\
Initial translational kinetic energy/eV: & 0.495 & &  \\
Distance for flux determination: &  $r$=12.0\AA &  &\\
Number of Chebychev iterations: &201000 for $J=0$ & 185000 for $J=10-60$ & 5000 for $J > 60$\\
\hline
\end{tabular}
\end{center}
\end{table*}

\begin{figure}[h]

\begin{center}
\includegraphics[width=9.cm]{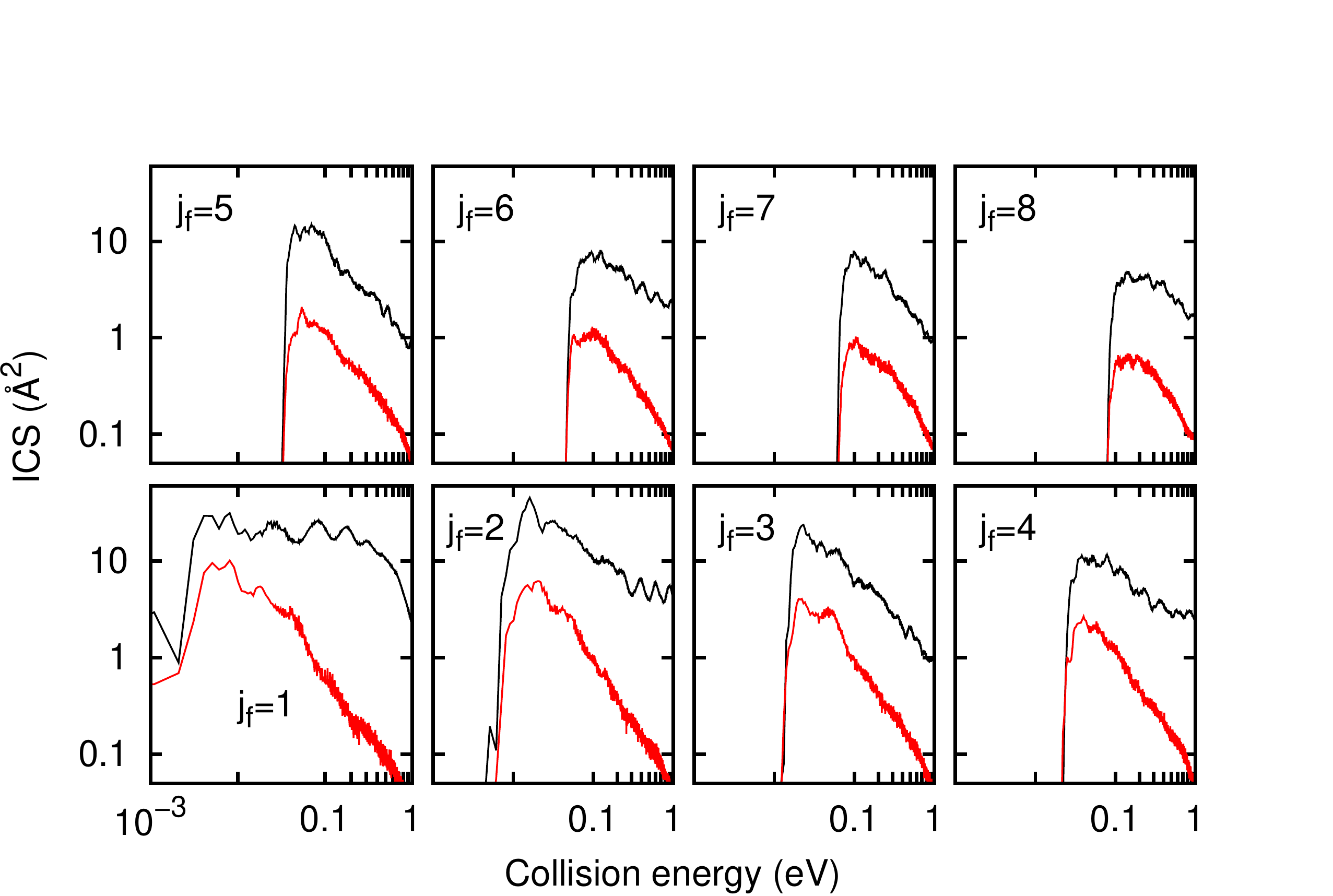}

\caption{\label{WP-doublet}{
    Cross section for the inelastic (solid, black) and exchange (dashed red) in the 
    SH$^+(^3\Sigma^-,v=0,j=0)$ + H collisions,
    calculated with the WP method on the doublet state.
}}
\end{center}
\end{figure}

The calculated cross sections for the SH$^+$(v=0,j=0) + H in the doublet state
in the inelastic and exchange channels are shown in Fig.~\ref{WP-doublet}. The exchange
cross section is more than one order of magnitude lower than the inelastic one. The doublet
state shows a deep well and as a consequence, the calculated reaction probabilities at all
the partial waves show the presence of many resonances. This necessitates propagation
of the wave packet for a high number of iterations. In spite of the presence of a
dense manifold
of resonance, the reaction is not statistical. The same situation was obtained for
OH$^+$+H dynamics in the doublet state \citep{Bulut-etal:15b},
since in the deep well the SH$^+$ distance does not vary with respect to that of bare SH$^+$,
and the light colliding H atom is not able produce a complete randomization of the available energy
between the two SH bonds, which would lead to a probability of 0.5 for each rearrangement channel.

\subsection{Formation and destruction rate constants}

Quantum rate constants are obtained from the corresponding cross sections, described above for
several reactive/inelastic collisions, by numerically integrating over collision energy with a Boltzmann distribution at each temperature and summing over the rotational state of the reactants with a weight determined from a Boltzmann distribution.
These rate constants are specific for each vibrational state of either H$_2$ or SH$^+$, for formation
and destruction processes, respectively.

Quasi-classical rate constants were also calculated  using the method of \cite{Karplus-etal:65}
as implemented in the miQCT code \citep{Zanchet-etal:13b,Dorta-Urra-etal:15,Zanchet-etal:16}.
For each temperature,
five hundred thousands  trajectories are run changing the initial conditions, consistent
with a Boltzmann distribution of translation and rotation energy at a given temperature $T$.
The initial distance between reactants is set to 21 \AA, with a maximum impact
parameter of 11.6 \AA. The trajectories are stopped when any distance becomes greater
than 24 \AA. { For the destruction reaction, SH$^+$
  is considered to be in $v=0$, and for the formation reaction, H$_2$ is 
  in $v=2$, 3, 4, and 5.}
{ A Boltzmann distribution for rotational excitation is considered in the QCT calculations
  of reaction rate for all initial vibrational excitations.}

\begin{figure}[h]

\begin{center}
\includegraphics[width=9.cm]{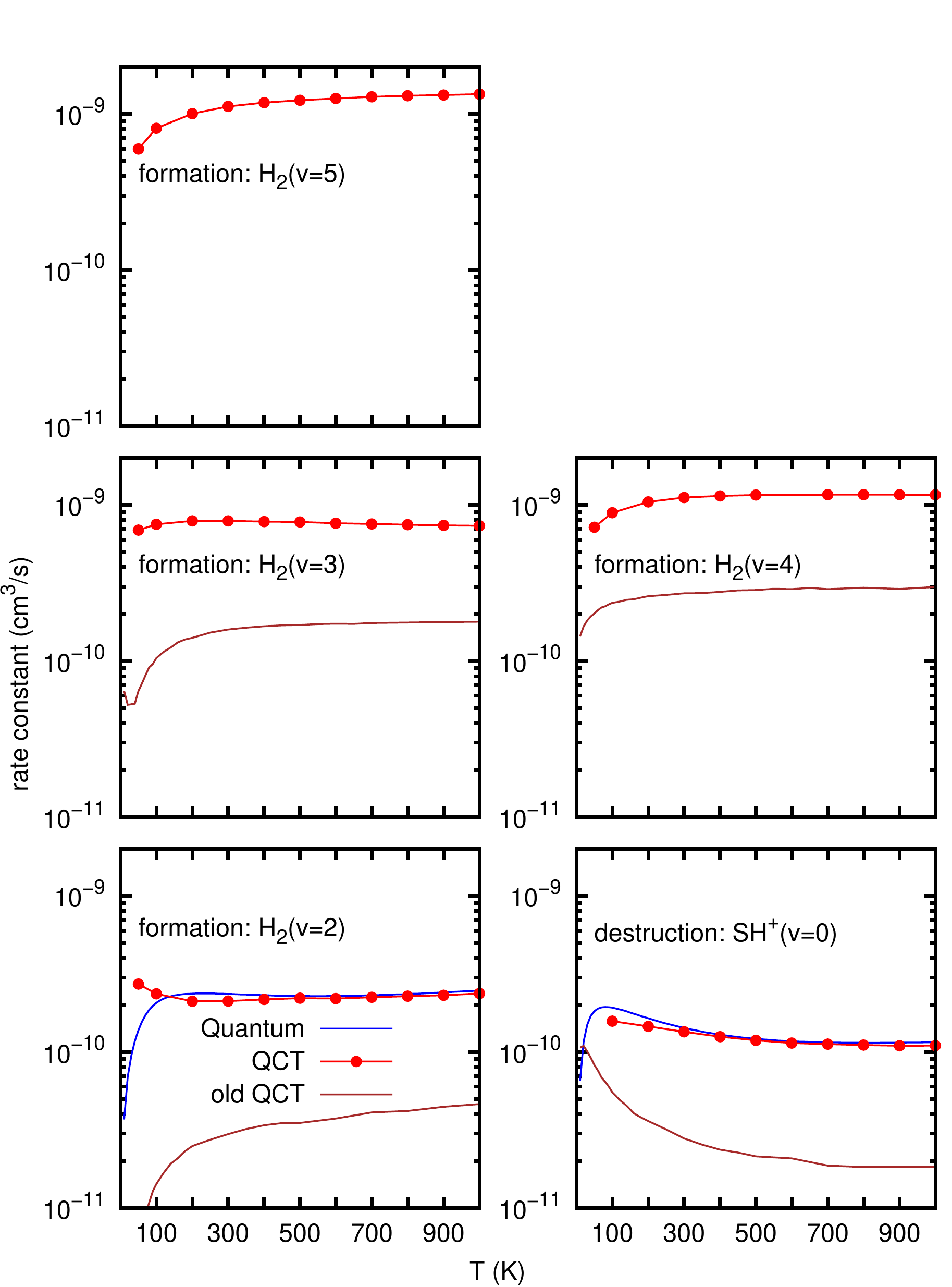}

\caption{\label{rates}{
    Quantum and classical thermal rate constants for the formation,
       S$^+$ + H$_2(v=2,3,4,5)$ $\rightarrow $ SH$^+$ + H,
       and destruction,
       SH$^+$(v=0) + H $\rightarrow $ S$^+$ + H$_2$. Also, the results obtained
        by \cite{Zanchet-etal:13b} are included for comparison.
       The destruction rates include the 2/3 electronic partition function contribution.
}}
\end{center}
\end{figure}

The calculated (quantum and QCT) reaction rate constants for formation and destruction processes
are shown in  Fig.~\ref{rates}, and compared to
those of \cite{Zanchet-etal:13b}, calculated using a QCT method.
The quantum and QCT calculations shown in the two lower panels
are in very good agreement for temperatures higher than 100-200 K. This indicates
that the QCT results for the rate constants are relatively accurate and justify the
use of this less time-consuming method to
calculate the formation rates for H$_2$($v=3$, 4 and 5).
Below 100 K, quantum effects are more important, and the quantum and classical
results diverge.

The QCT results obtained here are larger
than those reported from the previous quartet PES of \cite{Zanchet-etal:13b}. This difference
is attributed to the slight decrease in the endothermicity. 
The factor depends on the initial vibrational state, but we can conclude
that the present formation rate constants are  larger than those previously reported
by a factor of between three and six. The present QCT results
are also similar to those reported by \cite{Song-etal:18} using a different PES for the quartet.

The destruction rate constants of this work are also larger than  the previous ones reported
in \cite{Zanchet-etal:13b} by a factor that strongly depends on the temperature: about five at 1000 K  and  two at  100K.
The reason for this change is also attributed to the higher accuracy of the present PES.

Since the formation and destruction rate constants increase significantly with respect to those
reported previously \citep{Zanchet-etal:13b}, the predicted abundance of SH$^+$
in interstellar conditions should be influenced by these new data.
The formation and destruction quantum rate constants have been fitted to the usual expression
\begin{eqnarray}
  K(T)=\alpha\,\left({T\over 300}\right)^\beta\, \exp(-\gamma/T),
\end{eqnarray}
and the $\alpha$, $\beta$ and $\gamma$ parameters obtained are listed in Table~\ref{table-formation-destruction-rates}.

\begin{table}
  \caption{Parameters of the fit of the formation and destruction rate constants
     in the quartet electronic state\label{table-formation-destruction-rates}}
\centering
\begin{tabular}{cccc}
\hline \hline
\multicolumn{1}{c}{Reaction} & \multicolumn{1}{c}{$\alpha$} & \multicolumn{1}{c}{$\beta$} & \multicolumn{1}{c}{$\gamma$} \\
 & \multicolumn{1}{c}{(cm$^3$ s$^{-1}$)} & & \multicolumn{1}{c}{(K)}  \\
\hline
\multicolumn{4}{l}{Formation}\\
\hline
S$^+$ + H$_2$($v$=2)    $\rightarrow$ SH$^+$ + H   (*)   &     2.88 10$^{-10}$     & -0.15 &  42.93\\
S$^+$ + H$_2$($v$=3)    $\rightarrow$ SH$^+$ + H   \quad \quad  &     9.03 10$^{-10}$     & -0.11 &  26.15\\
S$^+$ + H$_2$($v$=4)    $\rightarrow$ SH$^+$ + H    \quad\quad  &     12.96 10$^{-10}$     & -0.04 &  40.80\\
S$^+$ + H$_2$($v$=5)    $\rightarrow$ SH$^+$ + H    \quad \quad  &     12.09 10$^{-10}$     & 0.09 &  34.51\\
\hline
\multicolumn{4}{l}{Destruction}\\
\hline
SH$^+$($v$=0) + H $\rightarrow$ S$^+$  + H$_2$   (*) &     1.86 10$^{-10}$     &  -0.41 & 27.38 \\
\hline
\end{tabular}
{Rate coefficient is given by $\alpha$ ($T$/300)$^\beta$ $\exp(-\gamma/T)$.}\hfill\\
{* fit to the quantum results\hfill}\\
\end{table}

\section{Updated predictions of the SH$^+$ abundance in the Orion Bar PDR}

We use the new 
\mbox{S$^+$ + H$_2$ $\rightleftarrows$ SH$^+$ + H}  rate constants to update the estimated 
abundance of SH$^+$ in a dense PDR like the Orion Bar \citep[][and references therein]{Goicoechea-etal:16}. 
The external and most UV-irradiated layers
of dense PDRs have moderate column densities of vibrationally excited H$_2$
that depend on the gas density and flux of UV photons. These H$_2$($v \geq 1$) molecules greatly
affect the formation of hydrides 
\citep[e.g., see models of][]{Sternberg:1995,Agundez-etal:10}.  
SH$^+$ has been detected in the Orion Bar
with Herschel \citep{Nagy-etal:13} and also from the ground \citep{Muller-etal:14}. \cite{Zanchet-etal:13b} have previously shown PDR models that emphasized the role
of  H$_2\,(v\geq2$) in the formation of SH$^+$. Indeed, ALMA images of the Orion Bar
\citep{Goicoechea-etal:17} have shown that the SH$^+$ emission comes from a narrow gas layer
that delineates the irradiated edge of the PDR. The SH$^+$ column densities estimated
from observations however,
revealed values that are higher than model
predictions by a factor of between 3 and 30 (depending on cloud geometry considerations).
The latter predictions were
made using  the Meudon PDR model
\citep[e.g.,][]{Lepetit-etal:06,Goicoechea:07}  adapted to the physical and chemical structure  of the Orion Bar.

\begin{figure}[h]

\includegraphics[width=9.cm]{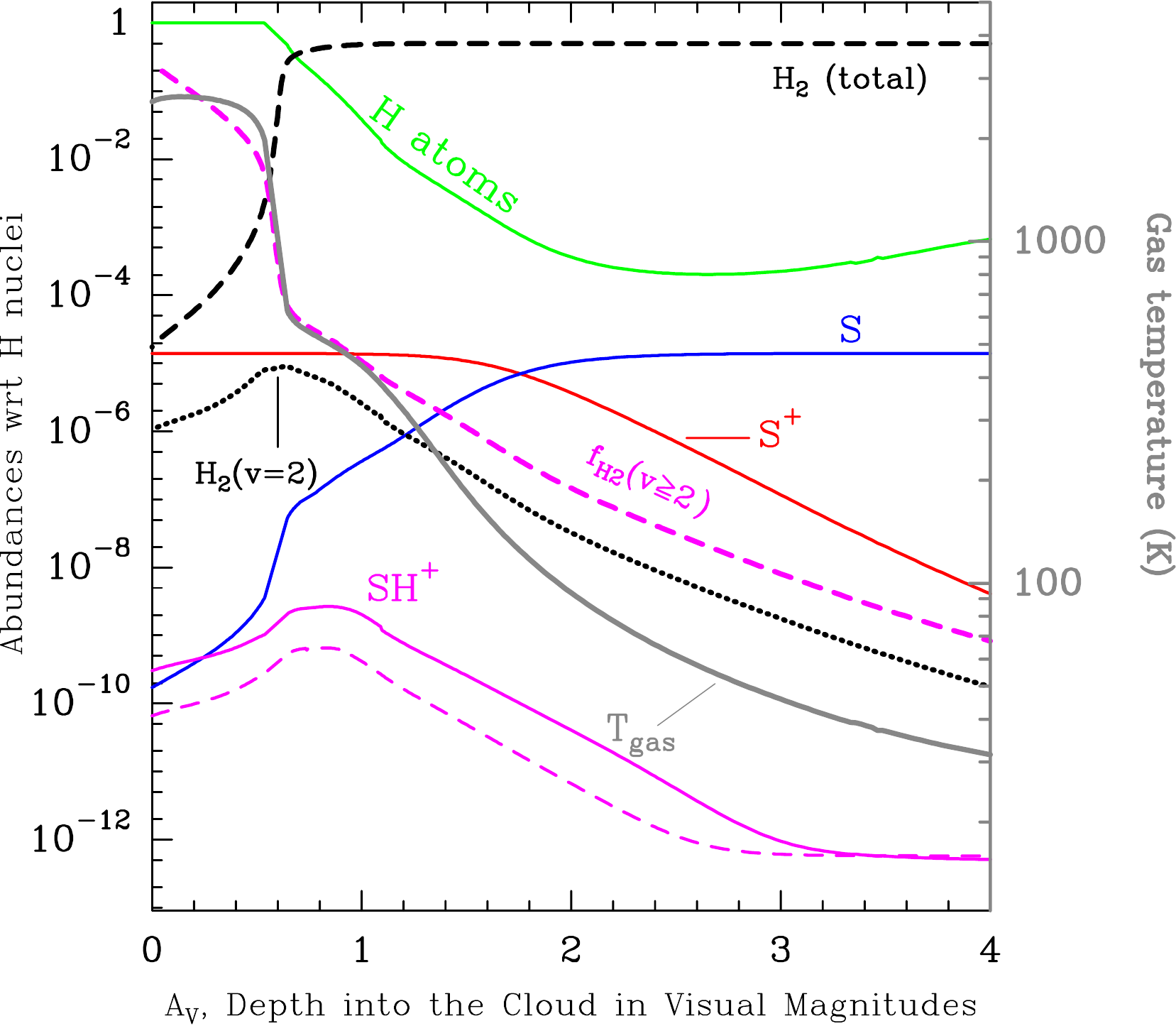}
\caption{Model of a strongly irradiated and constant-pressure PDR with
 \mbox{$P_{\rm th}$= 2$\cdot$10$^8$ K cm$^{-3}$} and  \mbox{$G_0$=2$\cdot$10$^4$}, appropriate to the most irradiated layers of the Orion Bar PDR. Fractional abundances of SH$^+$, \mbox{H$_2$ (total)}, \mbox{H$_2$ ($v$=2)}, S$^+$, S and, H are shown as a function of depth into the cloud. 
We also show $f_{\rm H_2}$, the fraction of H$_2$ that is in vibrationally excited
levels $v\geq2$, and the gas temperature (in gray, right axis scale). 
The dashed SH$^+$ abundance profile is for a model that
uses the old formation and destruction rates \citep{Zanchet-etal:13b}.
 \label{PDRmods}}
\end{figure}

The Orion Bar is illuminated by a far-UV ($<$13.6\,eV) radiation field $G_0$ 
of a few times 10$^4$, where $G_0 = 1$ is equal to 1.6$\times$10$^{3}$\,erg\,cm$^{-2}$\,s$^{-1}$, the far-UV flux in the solar neighborhood, integrated from $\sim$912\,\AA~to $\sim$2400\,\AA\  \citep{Habing:68}. Here we
used version 1.5.2 of the Meudon PDR code to model an isobaric PDR, with a constant gas thermal pressure of \mbox{$P_{\rm th} =n_{\rm H}$\, $T$\,=\,2$\cdot$10$^8$~cm$^{-3}$\,K} determined from previous observations \citep[see e.g.,][]{Goicoechea-etal:16,Joblin-etal:18}
and introduce the new SH$^+$ formation and destruction rate constants.
We adopt an impinging  FUV field of  $G_0 = 2 \cdot 10^4$,
an undepleted sulfur abundance of 1.4$\times$10$^{-5}$ with respect to H nuclei \citep{Asplund:05}, and dust grain properties appropriate to the flatter extinction
curve observed toward Orion \citep[i.e., an extinction-to-color-index ratio
$R_V = A_V / E_{B-V} = 5.5$, e.g.,][]{Cardelli:89}.
Figure \ref{PDRmods} shows fractional abundances of 
SH$^+$, \mbox{H$_2$ (total)}, \mbox{H$_2$ ($v=2$)}, S$^+$, and S, and atomic
hydrogen as a function of depth into the cloud (in magnitudes of visual
extinction $A_{V}$). The figure also shows the gas temperature (in grey), and $f_{\rm H_2}$, the fraction of H$_2$ that is in vibrationally excited
levels $v\geq2$ with respect to the ground 
\mbox{$f_{\rm H_2}$\,=\,$n\,(v\geq2)$/$n\,(v=0)$}. The SH$^+$ abundance profiles obtained from the older \mbox{S$^+$ + H$_2$ $\rightleftarrows$ SH$^+$ + H} rate constants of  \cite{Zanchet-etal:13b} (dashed curve) are also shown. 

As observationally shown by
\citet{Goicoechea-etal:17}, SH$^+$ is only abundant
in the irradiated edge of the PDR, with the abundance peak located at
$A_{V} \approx 1$.  The models that use the new formation and destruction rate constants produce  SH$^+$ column densities \citep[$N$(SH$^+$)$=$(0.2-2.7)$\cdot$10$^{13}$~cm$^{-2}$,
depending on the assumed inclination of the PDR, see][]{Goicoechea-etal:17}
that are higher 
than models using previous data  
 \citep{Nagy-etal:13,Zanchet-etal:13b,Goicoechea-etal:17}  by a factor of approximately four.
 This enhancement in the production of SH$^+$ reconciles the SH$^+$ column densities inferred
 from observations  with
 the PDR model predictions.  
These results further empathize the role of UV-pumped H$_2$ molecules  in the formation of simple molecules at the edge of strongly irradiated dense PDRs.

\section{Conclusions} \label{sec:conclusions}

New PESs for the $^4A''$ and $^2A''$ electronic states have been calculated
for the \mbox{S$^+$ + H$_2$ $\rightleftarrows$ SH$^+$ + H} reactions.
Quantum state-to-state reactive and inelastic cross sections have been calculated
for S$^+$ + H$_2$(v=2) and SH$^+$(v=0) + H cases, and the corresponding rate constants
were calculated and compared to QCT calculations. Good agreement was found.
This justifies the use of the QCT method for the calculation of the formation rate constants
for higher vibrational states, S$^+$ + H$_2$($v=3$, 4 and 5 ).
These new rate constants are larger than
those previously calculated \citep{Zanchet-etal:13b} by roughly a factor of between two and six.

We used the new reaction rates in the Meudon PDR code and simulated the UV-illuminating
and physical conditions in the Orion Bar PDR. The new models yield 
column densities that are a factor of four higher than those obtained with the previous formation and destruction rate constants \citep{Zanchet-etal:13b}.
The new fractional abundances of SH$^+$ are within the uncertainties of 
the SH$^+$ column densities inferred
 from observations \citep{Goicoechea-etal:17}. 

The spin-rotation couplings of 
SH$^+$($^3\Sigma^-$)  have not been included, and therefore the present state-to-state
rates cannot be directly used to model the fine structure observed for SH$^+$.
This is being done in a separate study
using a recoupling technique \citep{Faure-Lique:12}, in which it is assumed
that the electronic spin is a spectator during the collision.

\begin{acknowledgements}
We thank E. Bron for his help with the new version of the 
Meudon PDR code.  We acknowledge the French-Spanish collaborative project PICS (Ref. PIC2017FR7). 
 The research leading to these results has received funding from
 MICIU under grants No. FIS2017-83473-C2 and AYA2017-85111-P.
 FL acknowledges financial support from the Institut Universitaire de France.
 NB acknowledges the computing facilities by TUBITAK-TRUBA.
\end{acknowledgements}



\end{document}